\documentclass[prl,twocolumn,amssymb]{revtex4}
\usepackage{graphicx}
\usepackage{color}
\usepackage{epsfig}
\usepackage{latexsym}
\usepackage{bm}
\usepackage{ulem}
\usepackage{color}
\usepackage{color,graphicx}

\def\clr{\color{red}}

\begin{document}

\title{Continuously Varying Critical Exponents Beyond Weak Universality}

\author{N. Khan$^1$, P. Sarkar$^2$,  A. Midya$^1$ , P. Mandal$^1$ and P. K. Mohanty$^{1}$}\email{ pk.mohanty@saha.ac.in}

\affiliation{$^1$CMP Division, Saha Institute of Nuclear Physics,
1/AF Bidhan Nagar, Kolkata 700064, India \\ $^2$Department of
Physics, Serampore College, Serampore 712201 India}


\begin{abstract}

{\bf Renormalization group theory  does not
restrict the from  of    continuous variation of critical exponents
which occurs  in presence of a marginal operator.  However, the
continuous variation of critical exponents, observed in different
contexts, usually follows a weak  universality  scenario   where some
of the exponents (e.g., $\beta, \gamma, \nu$)  vary keeping others
(e.g., $\delta , \eta$) fixed.   Here we  report a ferromagnetic phase 
transition  in (Sm$_{1-y}$Nd$_{y}$)$_{0.52}$Sr$_{0.48}$MnO$_3$ $(0.5\le y\le1)$ 
single crystal where {\it all} critical exponents vary  with $y.$
Such variation clearly violates both  universality and weak  universality
hypothesis. We propose a  new scaling   theory that
explains the present experimental results, reduces to  the weak
universality as  a special  case,   and provides a generic route
leading  to continuous variation  of  critical exponents and
multicriticality.}
\end{abstract}

\maketitle

Study of critical phenomena is based on two concepts: one is
universality \cite{univ_hypo,univ_hypo1} which  states that the
associated critical exponents  and scaling functions  are  universal
up to symmetries and  space dimensionality, and another is scaling
theory \cite{KadanoffBook} that describes    the general properties
of the scaling  functions  and relates different critical exponents.
In the renormalization group approach \cite{Wilson}, the  critical
point is a fixed point  governed by   a unique set  of  relevant
operators with  scaling dimensions (critical exponents) which are
fully independent of irrelevant operators. While a relevant
perturbation may take the system to a new fixed point, the marginal
one  brings in a possibility of a  line of  critical points  having
continuous variation of exponents. Although  the  concept of
universality  has been verified experimentally time and again,
starting from early 40s \cite{Guggenheim} to present  \cite{recent},
a  continuous variation is rarely observed. A  clear  example   is
provided by Baxter \cite{Baxter} who solved the eight vertex model
\cite{sutherland} (EVM) exactly, and Kadanoff and Wegner
\cite{kadanoff}  who provided a mapping of EVM to a two-layer Ising
system with a marginal four-body interaction between the layers
\cite{Wu} (similar to Ashkin Teller model  \cite{AT,AT1,AT2}) that
drives the  continuous variation. In later  years, Suzuki
\cite{suzuki} proposed a {\it weak} universality (WU) scenario where
critical exponents (like $\beta,  \gamma,  \nu$   in EVM) may change
continuously but their ratios ($\frac \beta \nu, \frac \gamma \nu$
and consequently $\delta=1+\frac{\gamma}{\beta}$) remain invariant.
This  WU scenario has  been observed in   frustrated spin systems
\cite{AntiIsing2D, ASen}, interacting dimers \cite{dimers}, magnetic
hard squares \cite{MagneticHardsquare}, Blume-Capel models
\cite{Blume-Capel}, reaction diffusion systems
\cite{ReactionDiffusion}, absorbing phase transitions
\cite{Absorbing}, percolation models \cite{Percolation,
Percolation1}, fractal structures   \cite{fractal}, quantum critical
points \cite{QCP_Suzuki}, etc. The  generic   nature of the marginal
interaction that  leads  to weak universality in all these different
systems remain  unclear.

To  the best of our knowledge,  most systems which  show  continuous
variation  of  critical exponents obey weak universality  - a few
exceptions include criticality  in  Ising spin glass
\cite{IsingSpinGlass}, micellar solutions
\cite{MicellarSolutions,MicellarSolutions1}, frustrated spin systems
 \cite{KagomeAF}, strong coupling QED  \cite{Kondo} etc.
Unconventional exponents are also experimentally observed in several
systems like weak ferromagnet BaIrO$_3$ ($\beta \approx$0.82,
$\gamma \approx$1.03 and $\delta \approx$2.20), organic ferromagnet
TDAE-C$_{60}$ ($\beta \approx$0.75, $\gamma \approx$1.22 and $\delta
\approx$2.28), quasi-two-dimensional organic conductor
$\kappa$-(BEDT-TTF)$_2$Cu[N(CN)$_2$]Cl ($\beta \approx$1, $\gamma
\approx$1 and $\delta \approx$2), etc  \cite{bair03,tdae,nature}.
Recently,  Fuch et. al.  \cite{fuch} have  observed  the linear
variation of exponents in the polycrystalline samples of
Sr$_{1-z}$Ca$_{z}$RuO$_3$ from $\{ \beta \approx 0.5, \gamma \approx
1, \delta \approx 3\}$ for $z=0$ to $\{ \beta \approx$1, $\gamma
\approx0.9$ and $\delta \approx 1.6 \}$ for $z$=0.6,   which satisfy 
Widom scaling relation ($\beta+\gamma= \beta\delta$)  for  all $z.$ Authors have
suggested that the evolution of exponents may be originating from
orthorhombic distortions or additional quantum fluctuations
associated with quantum phase transition. 

Anomalous  ferromagnetic (FM)
transition has also been observed in
mixed valance manganites, $RE_{1-x}AE_x$MnO$_3$ ($RE$: rare earth
ions, $AE$: alkaline earth ions) either as a discontinuous
transition or a continuous transition with a set of critical
exponents that does not belong to any known universality or the weak
universality
\cite{review-manganites,QD-firstorder,PCMO-firstorder,LCMO-firstorder,
 QD-SSMO,demko,prb-09,prl-09,ESMO-firstorder,unusual-exponent-LCMO,unusual-exponent,Salamon}.
  The unusual magneto-electronic properties in narrow band  manganites are closely associated  with 
 the quenched disorder arising mainly due to the size mismatch between  
 $A$-site cations \cite{dagotto,rod}. Such disorder  
 reduces the carrier mobility and the formation energy for lattice polarons \cite{sato}, in effect $T_C$ reduces,  rendering   the FM  
 transition towards first-order.  
 A system  with  narrow  bandwidth  and large disorder such  as 
 $Sm_{1-x}Sr_x$MnO$_3$ shows a sharp
first-order FM transition for $x = 0.45 - 0.48$
\cite{QD-SSMO,demko,prb-09,prl-09}. The first-order transition is
however extremely sensitive to external pressure, magnetic field,
$A$-/$B$-site substitution, oxygen isotope exchange, etc. - with the
application of external and internal pressure (chemical
substitution) beyond a critical threshold, the transition becomes
continuous \cite{demko,prb-09,prl-09,O2-isotope}.

In this  paper, we report two important results - a first-time
observation  of a thermodynamic transition ({\it i.e.} FM phase
transition in (Sm$_{1-y}$Nd$_y$)$_{0.52}$Sr$_{0.48}$MnO$_3$) where
{\it all}  critical exponents  vary continuously, and  a  new scaling
theory that  unifies the earlier observations of  continuous
variation of exponents along with the  present experimental results.
We  propose that,  to  obey the  scaling relations  consistently,
the  variation of critical  exponents are constrained to   have
specific forms.  This scaling   hypothesis  naturally leads to two
special  cases which have been  realized  earlier, namely the weak
universality \cite{suzuki} (where $\delta$ is fixed)  and the strong
coupling QED \cite{Kondo} (fixed $\gamma$). A more   generic
scenario is   the one  which  allows simultaneous variation  of {\it
all} the critical exponents in an intrigue way, leading to  a
multi-critical point where  the phase transition becomes
discontinuous. This scenario is verified  experimentally in a
comprehensive and systematic study of FM phase transition in
(Sm$_{1-y}$Nd$_y$)$_{0.52}$Sr$_{0.48}$MnO$_3$ single crystal. For
higher doping concentration $y>0.4,$ the  FM transition is found to
be  continuous,   but to our surprise,  the critical exponents
exhibit continuous variation with Nd concentration $y,$ starting
from  $(\beta, \gamma, \delta) = ( 0.16, 1.27, 9.30)$  at $y=0.5$ to
$(0.36,1.38,4.72) $ at $y=1.$ Within error limits,  $y=1$  belongs
to the universality class of Heisenberg model in three dimension
(HM3d). The proposed scaling hypothesis successfully explains the
continuous variation  of exponents and predicts  a discontinuous
transition for $y < 0.37$ which  has been observed  recently  by
others \cite{demko, prb-09}.

\section{Experimental Methods}

The single crystals of (Sm$_{1-y}$Nd$_y$)$_{0.52}$Sr$_{0.48}$MnO$_3$
with $y=0.5, 0.6, 0.8$ and $1.0$  were prepared by floating zone
technique under oxygen atmosphere \cite{Prabhat-da,Prabhat-da1}.
Single crystallinity was confirmed by the Laue diffraction.  The dc
magnetization measurements were performed using a Quantum Design
magnetic property measurement system (MPMS SQUID VSM) in fields up
to $7$ T. The data were collected after stabilizing the temperature
for about $30$ min. External magnetic field was applied along the
longest sample direction and data were corrected for the
demagnetization effect.

\section{Results and Discussion}

\subsection{Critical temperature and exponents :}  
\begin{figure}[ht]
\centering
\includegraphics[width=\linewidth]{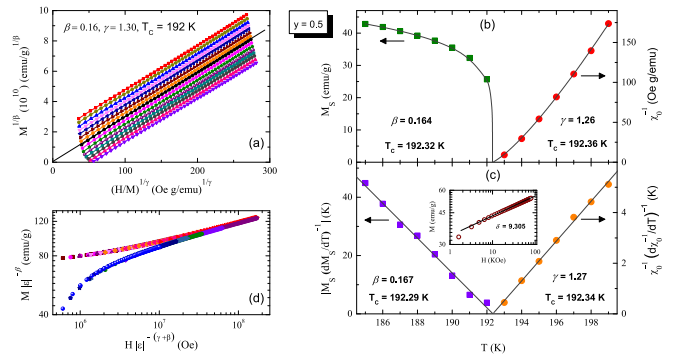}
\caption{(a) Modified Arrott plot [$M^{1/\beta}$ vs
$(H/M)^{1/\gamma}$] isotherms (185 K$\leq T \leq$199 K in 1 K
interval) of (Sm$_{1-y}$Nd$_{y}$)$_{0.52}$Sr$_{0.48}$MnO$_3$
($y$=0.5) single crystal. Solid lines are the high-field linear fit
to the isotherms. The isotherm (at $T$ = 192 K) closest to the Curie
temperature ($T_C$ = 192.3 K) almost passes through the origin in
this plot. (b) Temperature dependence of spontaneous magnetization,
$M_S$ (square) and inverse initial susceptibility, $\chi_0^{-1}$
(circle). Solid lines are the best-fit curves. (c) Kouvel-Fisher
plots of $M_S$ and ${\chi_0}^{-1}$. Inset shows log scale plot of
M(H) isotherm at $T = T_C$ (d)  Scaling  collapse of $M-H$ curves
following  Eq. (\ref{eq:scaling11}), indicating two universal curves
below and above $T_C.$} \label{fig:Arrott}
\end{figure}

Let  us set
notations by reminding that in  absence of magnetic field ($H$) the
spontaneous magnetization  of the system  vanishes as   $M_S (0,
\varepsilon) \sim (-\varepsilon)^\beta$  and  the initial
susceptibility diverges as $\chi_0 (0,\varepsilon) \sim
(\varepsilon)^{-\gamma}$ as the critical point is approached, {\it
i.e.,}  when  $(T/T_C-1)\equiv \varepsilon \to 0.$ Again at $T =
T_C,$ the magnetization varies algebraically $M (H,T_C) \sim
H^{1/\delta}$  \cite{cstanley}.  To estimate the critical exponents
$\beta,\gamma$ and $\delta,$ we need to   know $T_C$ accurately. To
do so, we exploit the linearity   in Arrott-Noaks equation of state
\cite{arrott}
\begin{equation}\label{map}
\left(\frac{H}{M}\right)^{1/\gamma}=a  \varepsilon   +b M^{1/\beta},
\end{equation}
where $a,b$ are non-universal constants. The correct  choice of
$\beta$ and $\gamma$  can make the isotherms of $M^{1/\beta}$ versus
$(H/M)^{1/\gamma}$  a set of parallel straight lines  with  {\it
one}  unique critical isotherm  that  passes  through  the origin.
This is explained in Fig.  \ref{fig:Arrott}(a) for $y=0.5$ and the
self consistency is achieved for the values $\beta=0.16$,
$\gamma=1.30.$ The  isotherm   $T=192$ K passes almost  through the
origin. The  Arrotts  plots   can  be used further to get  better
estimates of $\beta,\gamma$ and $T_C.$ From the intercepts of these
parallel lines on $M^{1/\beta}$ and $(H/M)^{1/\gamma}$ axes, we
obtain $M_S$ and $\chi_0^{-1}$ for different temperature which are
shown in Fig.  \ref{fig:Arrott} (b).
 The  best  power-law fit  gives rise to  $\beta=0.164\pm 0.002$ with
$T_C=192.32\pm0.04$ K and $\gamma=1.260\pm0.003$ with
$T_C=192.36\pm0.03$ K, respectively.  As a final  estimate, we take
$T_C= 192.3$ K, $\beta=0.16$ and   $\gamma=1.26.$ These estimates
are in fact consistent with Kouvel-Fisher   criteria   \cite{ckf}
which predict  that in the scaling regime, both $M_S
\left(\frac{dM_S}{dT}\right)^{-1}$ and $\chi_0^{-1}
\left(\frac{d\chi_0^{-1}}{dT}\right)^{-1}$ are proportional to  $T$
with proportionality constants  $\beta^{-1}$ and  $\gamma^{-1}$
respectively; which is shown  in Fig. \ref{fig:Arrott}(c). To
estimate another critical  exponent $\delta,$  we  take the help of
$M$-$H$ isotherm at  $T\simeq T_C$ (here  $T=192$K).   The log-scale
plot  of $M$ vs. $H$, as   shown in   the inset   of  Fig.
\ref{fig:Arrott}(c), is linear  with slope $\delta=9.30.$
Following the same method, we have determined the critical exponents
and $T_C$s' for   $y = 0.6, 0.8$ and $1.0$ \cite{SI} which are
listed in Table \ref{table} along with the  critical exponents of
HM3d.  {\clr The  variation  of $T_C$ with $y$ can  be explained in terms of 
bandwidth and  the  local  disorder \cite{rod} of  the system.  With  increasing $y$,  the bandwidth  
increases  whereas  disorder  decreases; both  these effects  enhance   $T_C.$ }
It is   clear  from Table \ref{table}  that the  exponents  for $y=1$ are same as that of HM3d within
error limits whereas  they  deviate substantially and vary
systematically  when  $y$ is decreased. 
 Clearly,  $\beta$   decreases   whereas $\delta$ increases  by {\it two}-fold  as   one  goes from $y=1$  
to $y=0.5.$ On the other hand, for each $y=0.5, 0.6,0.8,1,$ the Widom scaling  relation $\beta+\gamma= \beta\delta$  is  satisfied.
This  indicates that  the   change  in $\gamma$  is  minimal, as  observed here.
The variation of  critical
exponents with respect to a system  parameter contradicts
universality,  moreover it violates   WU  as $\delta$  varies along
with $\beta$ and $\gamma.$ The   focus of current work  is  to
address this issue in details, but  first let us ask  whether the
scaling hypothesis, crucial for the description of near critical
systems, is valid here.

\begin{table}[ht]
\centering
\begin{tabular}{|l|l|l|l|l|}
\hline
$y$               & $T_C$ (K)            &$\beta$     &$\gamma$     &$\delta$    \\
\hline
0.5               & 192.3                & 0.16       & 1.27        & 9.30        \\
\hline
0.6        & 222.5                & 0.23 & 1.30 & 6.31  \\
\hline 0.8                             & 241.3                & 0.31
& 1.32        & 5.14 \\
\hline 1.0                             & 265.3 & 0.36 & 1.38
& 4.72 \\
\hline
HM3d &- &0.365  &1.386 & 4.82 \\
\hline
\end{tabular}
\caption{\label{table} Critical exponents of
(Sm$_{1-y}$Nd$_{y}$)$_{0.52}$Sr$_{0.48}$MnO$_3$}
\end{table}

For a thermodynamic  system   near FM transition, magnetization $M$
depends on   $H$ and $\varepsilon$  and follows a universal scaling
form  \cite{cstanley}
\begin{equation}\label{eq:scaling11}
M(H,\varepsilon) = |\varepsilon|^{\beta} {\cal
F}_\pm(H/|\varepsilon|^{\beta +\gamma}),
\end{equation}
where ${\cal F}_{+}$ is for $T > T_C$ and ${\cal F}_{-}$ is for $T <
T_C$. The utility of universal scaling function  lies in the fact
that   the $M$-$H$ curves obtained for different   $T$ (near $T_C$)
can be collapsed on to a single curve  when one plots $M/
|\varepsilon|^{\beta}$ versus $H/|\varepsilon|^{\beta +\gamma}.$
This scaling collapse  is shown in Fig. \ref{fig:Arrott}(d) for
$y=0.5$ (and  in \cite{SI}  for other values of $y$); the two
branches  in this curve correspond to supper- and sub-critical
phases. For $y=0.6,0.8$ and $1.0$, the data collapse is also shown
in Fig. \ref{fig:collapse}(a), where  we  use an alternative but
equivalent form  of Eq. (\ref{eq:scaling11}),
\begin{equation}
 M(H,\varepsilon)^{-1/\beta} = \varepsilon
{\cal  F}(\varepsilon H^{-1/(\beta +\gamma)}).\label{eq:scaling1}
\end{equation}
It is advantageous   to use  this form  as  the two branches  of Eq.
(\ref{eq:scaling11}) are now merged  to a  single
function ${\cal  F}$  with its  argument $x=\varepsilon H^{-1/(\beta
+\gamma)}$  extending  from the sub-critical ($x<0$) to  the
supper-critical ($x>0$) regimes.  A very good   data collapse  and - which
confirms that the scaling hypothesis  is in place and the value of
$T_C$ and critical exponents obtained through several prescriptions
are unambiguous and self-consistent.

\section{The scaling  hypothesis } 

In the following,  we propose
a new scaling ansatz  which  explains the experimental findings
presented here. 
Since  diverging fluctuations  are  known  to be the origin of
power-laws \cite{pk, pk1}, we start with  a scaling relation  that  relates the  exponents 
of energy and magnetic fluctuations (i.e.,  $\alpha$ and  $\gamma$) with   that  of   
diverging correlation  length  associated with criticality: 
\begin{equation}
2-\alpha=    d \nu  =    \gamma \frac{\delta+1 }{\delta-1},
\label{eq:scaling}
\end{equation}
where   $\nu,\alpha$  are exponents associated with  the correlation length $\xi \sim \varepsilon^{-\nu}$
and  specific heat $C_v \sim \varepsilon^{-\alpha}$,  $d$ is the dimension. The  first
equality  here relates  the  diverging  correlation length to  the
energy  fluctuation whereas the second equality  relates the same to
the  magnetic fluctuation ($\chi_{_{AC}} \sim
\varepsilon^{-\gamma}$ and  $\chi_{_{DC} } \sim H^{1/\delta-1}$).
If   a set of exponents   originates from {\it
one} underlying   universality class then   they  must  relate
to  the exponents of the  parent universality
$\{\alpha_0, \nu_0, \gamma_0, \delta_0\}$ as
\begin{eqnarray}
\nu = \frac{\nu_0}{\lambda^{\kappa+\omega}}; 2-\alpha = \frac{2-\alpha_0}{\lambda^{\kappa+\omega}};\gamma =
\frac{\gamma_0}{\lambda^{\omega}};
\frac{\delta+1 }{\delta-1} =\frac{1}{\lambda^{\kappa}} \frac{\delta_0+1 }{\delta_0-1}, \label{eq:scale_main}
\end{eqnarray}
so that   the  universal scaling  relation  Eq.  (\ref{eq:scaling}) remains valid.
 Here,   $\lambda$ is the  marginal parameter that  drives  the  continuous variation and
$\omega$ and $\kappa$  are  parameters  yet to be determined.
In    Eq. (\ref{eq:scale_main}), variation of $\delta, \gamma$  are considered  independent
- equivalently one  may  vary   $\eta,\nu$ independently and deduce
variation of other  exponents from scaling relations \cite{SI}.

\begin{figure}
\includegraphics[width=.8\linewidth]{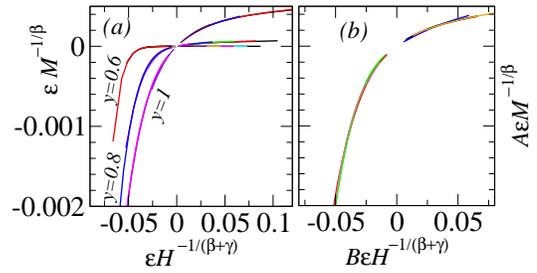}

\caption{(a) Scaling  collapse   of $M-H$ curves for $y=0.6,0.8$ and
$1$ following  Eq. (\ref{eq:scaling1}).  Although they appear
different, the   scaling functions  for $y=0.5,0.6,0.8,1$ can be
collapsed onto each other (shown in (b)) by rescaling of axis.}
\label{fig:collapse}
\end{figure}

Clearly $\kappa=0=\omega$ is  the parent universality  class
having  unscaled  ($\lambda=1$) exponents.  Other special cases  are   when  $\kappa$   or  $\omega$
vanishes  -then corresponding exponent $\delta$ or  $\gamma$  remains   fixed.
When $\kappa=0,$  the  continuous variation is governed  by  $c=\lambda^\omega.$
 In this case we may set  $\omega=1$ without loss of generality and
identify $c\equiv \lambda$   as the physical parameter that  drives  a continuous variation
\begin{eqnarray}
\nu =\frac{\nu_0}{\lambda};\gamma =\frac{\gamma_0}{\lambda} ;
2-\alpha = \frac{2-\alpha_0}{\lambda}; \beta =
\frac{2-\alpha-\gamma}2 =\frac{\beta_0}{\lambda}.
\end{eqnarray}
This  scenario  is already known as weak universality where
$\delta= 1+ \frac{\gamma} {\beta}=\delta_0$ and
$\eta  = 2-\frac{\gamma} {\nu} = \eta_0$ are universal as  in the old universality argument \cite{suzuki}.
Another special case  is $\omega=0$ and $\kappa$ set to unity,  where  $\gamma$   remains  invariant  and
\begin{eqnarray}
\nu =\frac{\nu_0}{\lambda};   \delta = \frac{(1+\lambda) \delta_0  +(1-\lambda)}{(1+\lambda)+(1-\lambda)\delta_0} ;
2-\alpha = \frac{2-\alpha_0}{\lambda}.\label{eq:QED}
\end{eqnarray}

This variation, with $\lambda$  being the   gauge coupling constant,
has been observed in strong coupling  QED  \cite{Kondo}.
Now we  turn  our attention to the  generic case   $\kappa>0$ and
set $\omega=1$ (generality is not compromised); the  consequent
variation of critical exponents is given by
\begin{eqnarray}
\nu&=& \frac{\nu_0}{\lambda^{\kappa+1}}~;~ \gamma =
\frac{\gamma_0}{\lambda}~ ;~ \delta= \frac
{(1+\lambda^{\kappa})\delta_0 +(1-\lambda^{\kappa}) } {
(1+\lambda^{\kappa}) + (1-\lambda^{\kappa})\delta_0} ; \cr \beta &=&
\frac{\beta_0  \delta_0 \gamma_0}{ \lambda \beta_0 (\delta
-\delta_0)  +\lambda \delta \gamma_0} ~;~ 2-\eta = \lambda^{\kappa}
(2-\eta_0) ~;~~\cr& &2-\alpha = \lambda^{-(\kappa+1)} (2-\alpha_0).
\label{eq:main}
\end{eqnarray}
Note, that  this   generic variation of exponents  satisfies 
Widom scaling relation $\delta = 1+ \gamma/\beta$  for  any $\lambda$ as long as  
parent universality  class  obeys the same.

We   further  include a  possibility that this  generic variation
may lead to  $\beta\to 0,$  a special    limit where  the phase
transition becomes discontinuous. At this  multi-critical limit the
usual scaling theory  demands $\gamma \to 1$   and   $\delta^{-1}\to
0$ \cite{re}. The first requirement  can   be met  if  the
multicritical  point occurs at $\lambda =\gamma_0$ and the second
requirement determines
\begin{equation}\label{kappa}
\kappa = \frac{1}{\ln \gamma_0}  \ln \frac{\delta_0+1}{\delta_0-1}.
\end{equation}

\begin{figure}
\centering
\includegraphics[width=.9\linewidth]{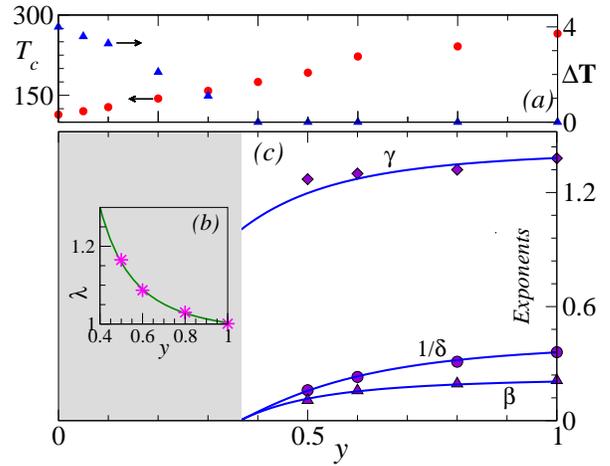}
\caption{(a) Critical temperature  $T_C$  for heating cycle (red circle)  and  the  thermal hysteresis 
width $\Delta T$ (blue triangle) for different Nd concentration $y.$ 
The values  for $y <0.5$ are taken from Ref.  \cite{prb-09}. (b) The proposed scaling
hypothesis has one free parameter $\lambda$ that maps to  $y$ -the
best choice gives $y(\lambda)=0.26/(\lambda - 0.97)^{0.4}$ (solid
line). (c) Critical exponents  for different $y$ from experiments
(symbol)  are compared with  Eq. (\ref{eq:main}) (solid lines).
Since $\beta\simeq 0$ at $y\simeq 0.37$, one expects a discontinuous
transition $y <0.37$ (also observed in (a)).} \label{fig:scale}
\end{figure}

In the  following, we   show that   FM   transition  in
(Sm$_{1-y}$Nd$_{y}$)$_{0.52}$Sr$_{0.48}$MnO$_3$ exhibits this new
universality  scenario. For the  present experimental system, the
parent  class is  HM3d (at $y=1$) and  accordingly $\kappa= 1.29$
(from  Eq. (\ref{kappa}) and Table \ref{table}). At  doping  $y=1$,
we may set  $\lambda=1,$  but a  general correspondence  between $y$
and  and the  marginal parameter $\lambda$ will not be known unless
we  know  the correct interaction Hamiltonian.  The   best choice of
$\lambda$  that    matches  the exponents  for $y=(0.5,0.6, 0.8, 1)$
well  turns out to be $(1.165, 1.087, 1.03,1.001).$ The  best fit to
this data as $y =A (\lambda-\lambda^*)^a$ results  in $A=0.26,
\lambda^*= 0.97$ and $a=-0.4$, which is shown in Fig.
\ref{fig:scale} (b). This is  used further   in Eq. (\ref{eq:main})
to get continuous variation of $\beta, \gamma$ and $1/\delta$ with
respect to $y$ (shown as solid line in Fig. \ref{fig:scale} (c)). A
good match here supports  the the new  scaling hypothesis strongly.
Moreover Eq. (\ref{eq:main}) suggests that  the transition becomes
discontinuous if $\lambda>\gamma_0 =1.386$  which corresponds to
$y\leq 0.37$.  In fact, a very sharp growth in the ordered moment
along with thermal  hysteresis   
just below $T_C$ has been   observed  in present system  for $y\le
0.4$ \cite{prb-09}, which is also  shown in Fig. \ref{fig:scale}(a).
In a related system (Sm$_{1-y}$Nd$_{y}$)$_{0.55}$Sr$_{0.45}$MnO$_3,$  where
Sr concentration differs slightly,   the 
transition  remained  first order upto  $y=0.33$ \cite{demko}.

To  emphasize that the new scaling  hypothesis  is indeed  in work,
we further investigate the scaling functions. If the critical
behaviour for different $y$ are related  to HM3d,  one expects that
the seemingly different universal scaling functions  ${\cal F}$  in
Fig. \ref{fig:collapse} (a) must collapse onto each other if  $x$-
and $y$-axis  are scaled  by  suitable constants. This  is described
in  Fig. \ref{fig:collapse} (b) - a good   collapse of scaling
functions for $y=0.5, 0.6,0.8, 1$ to a  unique universal form
further supports  that the observed criticality is only a rescaled
form of Heisenberg fixed point.

\section{Conclusions}

In conclusion, we have made a comprehensive study of critical
phenomena in single crystals of
(Sm$_{1-y}$Nd$_{y}$)$_{0.52}$Sr$_{0.48}$MnO$_3$ with 0.5$\leq y
\leq$1. The values of critical exponents $(\beta, \gamma,\delta)$
measured   for   $y=1$  are  consistent with Heisenberg universality
class  in  three  dimension,  whereas the same for $y=0.5, 0.6,0.8$
are far from any known universality class. All  these exponents vary
continuously   with $y,$ but  they seem to obey the standard scaling
laws following a single equation of state.  Variation of exponents
is not   new  to critical phenomena as it can be generated by a
marginal  interaction, but  most examples (though  there are a few
exceptions) in both theoretical and experimental  studies    satisfy
the  weak universality  \cite{suzuki} where   $\{\beta,\gamma,
\nu\}$  vary but $\{\eta, \delta\}$ are fixed. We argue that,  to
be  consistent with scaling,   the  continuous variation  must occur
in specific ways. Two  special cases are,   (a) $\delta$ remains
unchanged, which    leads  to  the weak universality, and (b)
$\gamma$ is unaltered, which results in  a  kind of variation
observed in strong coupling QED \cite{Kondo}. This generic
universality scenario, which leads to a multi-critical point,
explains the continuous variation of critical  exponents observed in
(Sm$_{1-y}$Nd$_{y}$)$_{0.52}$Sr$_{0.48}$MnO$_3$  and correctly
predicts the possibility of a discontinuous transition  for doping
$y<0.37.$

A  marginal interaction   that could provide  variation of  critical
exponents  beyond  weak  universality remains elusive - it is
certainly   challenging  to device  a microscopic theory to
accommodate  this  phenomenon.    Here,  
disorder   certainly  plays a role in  generating  anomalous   magnetic
fluctuations.
In addition,  
quantum fluctuations, if present,    may  bring in additional
features to the proposed scaling hypothesis; such a possibility has been
conjectured  in  recent experiments \cite{fuch},  but  a rigorous
and convincing  microscopic    theory is  still lacking. In
particular,  it is not clear, how  or why  a marginal  operator is
generated  in all the above experimental conditions to drive
continuous variations in specific  ways. \\

{\it Acknowledgements :} The authors would like to thank Mr. A. Pal for technical assistance.


%
%
%
%


\begin{thebibliography}{99}
\bibitem{pk1} A detailed  proof  for non-equilibrium systems  is  given in \cite{pk}, which  also 
holds trivially  for equilibrium.
\end{thebibliography}
\end{document}


\title{Supplementary Information : Continuously Varying Critical Exponents Beyond Weak Universality}

\author{N. Khan$^1$, P. Sarkar$^2$,  A. Midya$^1$ , P. Mandal$^1$ and P. K. Mohanty$^{1,*}$}
\affiliation{
$^1$CMP Division, Saha Institute of Nuclear Physics, 1/AF Bidhan Nagar, Kolkata 700064, India\\
$^2$Department of Physics, Serampore College, Serampore 712201,
India \\
$^*$Correspondence and requests for materials should be addressed to
P.K.M. (email: pk@saha.ac.in)}

\maketitle

In this supplement, we provide additional details  of  the
experiments and  the scaling theory. We  added two supplementary
figures that describe  the  measurement of critical exponents  on
(Sm$_{1-y}$Nd$_y$)$_{0.52}$Sr$_{0.48}$MnO$_3$ single crystals with
$y=0.6,0.8$  and 1. Also we  provide an  alternative   scaling
transformation of critical exponents. \\

\section*{Supplementary figures}

The supplementary  Fig. 1 describes determination of the critical
point  and exponents, and the  scaling collapse for Nd doping $y=1$-
within error limits they are same as the critical exponents  of
Heisenberg   Model  in three dimension. Figure 2 describes the same,
but for   different doping   $y=0.6$ and $0.8.$

\begin{figure}[ht]
\centering
\includegraphics[width=\linewidth]{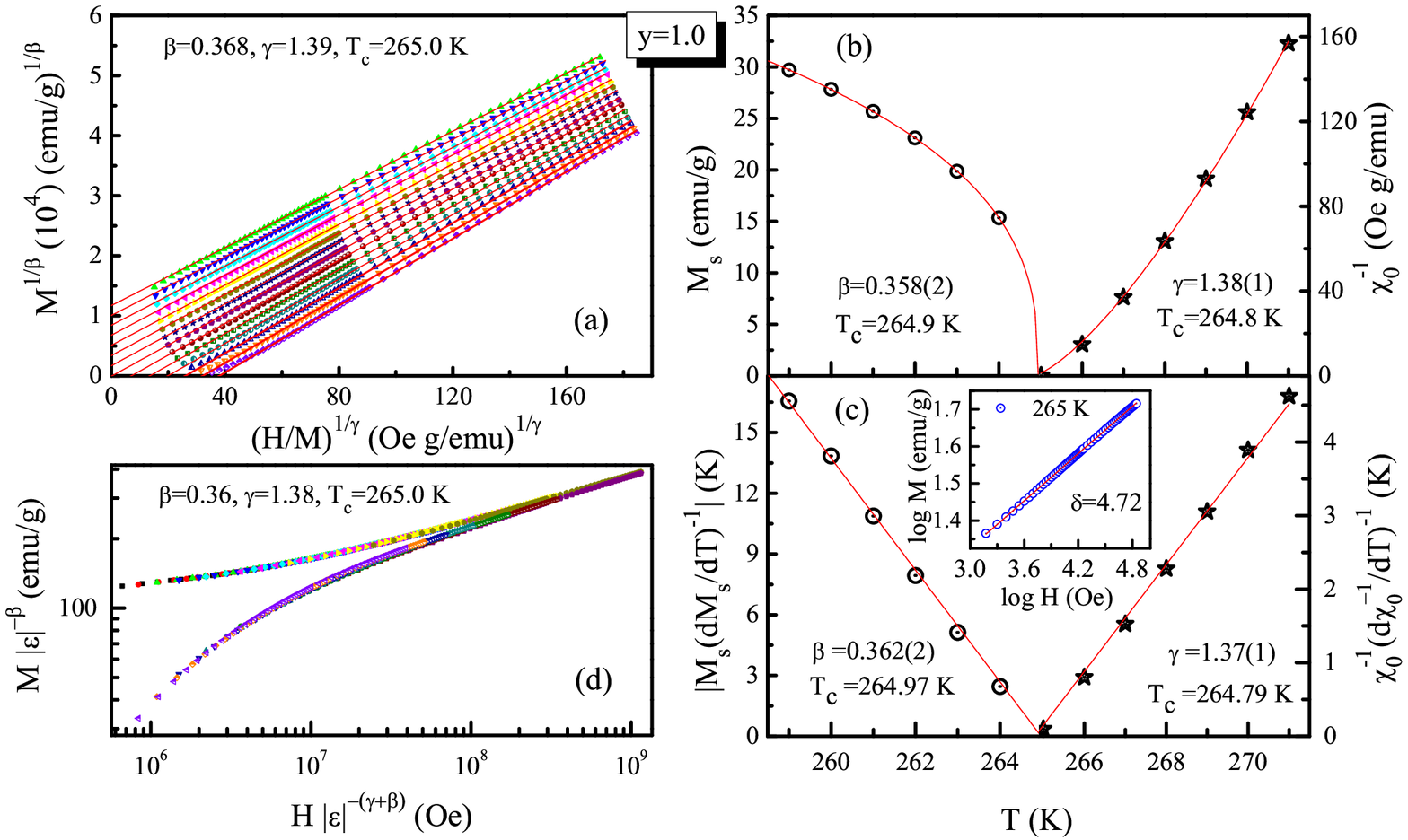}
\caption{(a) Modified Arrott plot [$M^{1/\beta}$ vs
$(H/M)^{1/\gamma}$] isotherms  ($258$ K $\le T \le  271$ K in $1$ K
interval) of (Sm$_{1-y}$Nd$_{y}$)$_{0.52}$Sr$_{0.48}$MnO$_3$ ($y$=1)
single crystal. Solid lines are the high-field linear fit to the
isotherms. The isotherm at $265.0$ K almost passes through the
origin in this plot. (b) Temperature dependence of spontaneous
magnetization, $M_S$ (circle) and inverse initial susceptibility,
$\chi^{-1}$ (star). Solid lines are the best-fit curves. (c)
Kouvel-Fisher plots of $M_S$ and ${\chi_0}^{-1}$. Inset shows
log-log plot of M(H) isotherm at $T = T_C$ (d)  Scaling  collapse of
$M-H$ curves: $M|\epsilon|^{-\beta}$  is an universal function of $H
|\epsilon|^{-(\beta+\gamma)},$ indicating two universal curves below
and above $T_C.$}
\end{figure}

\begin{figure}[ht]
\centering
\includegraphics[width=\linewidth]{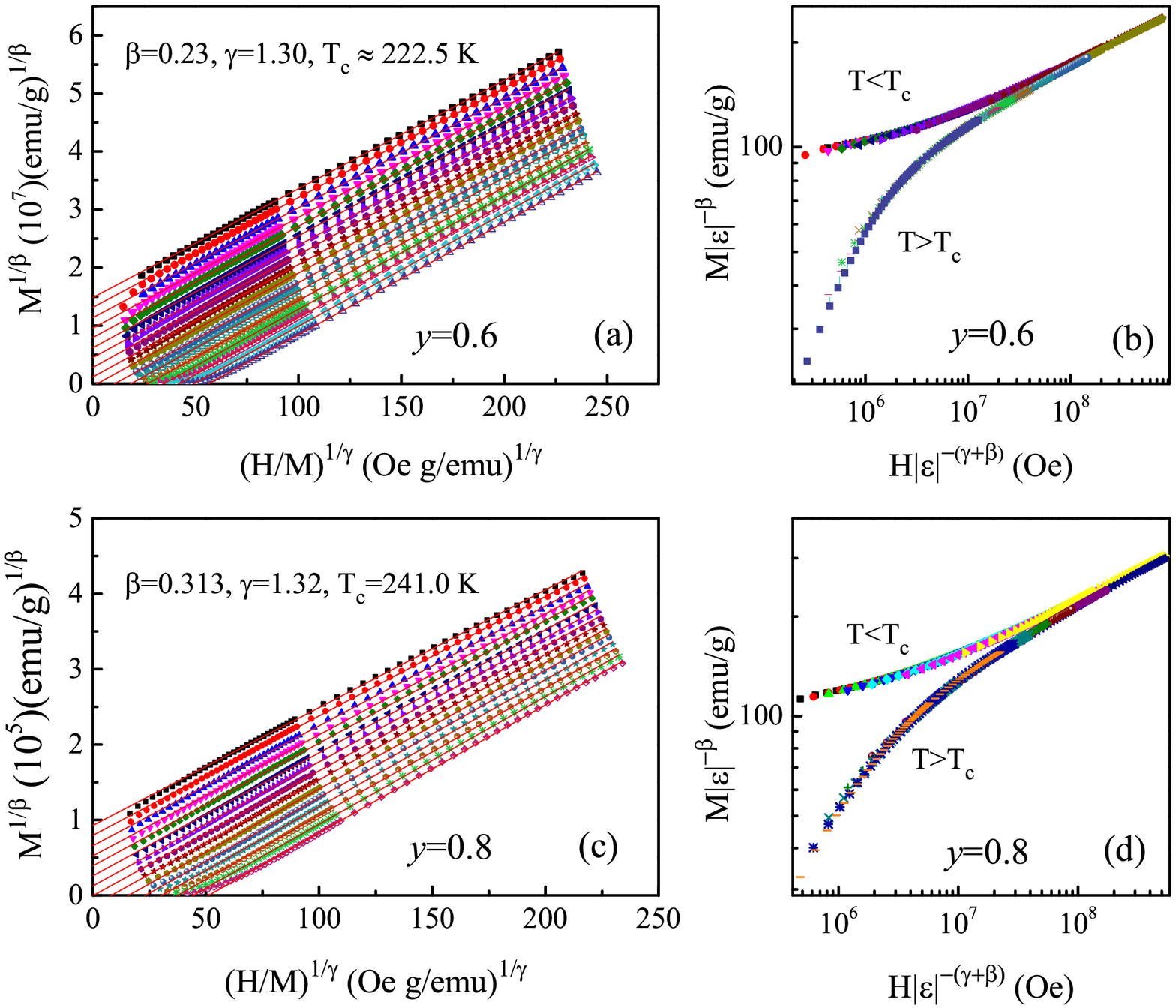}
\caption{(a) Modified Arrott plot [$M^{1/\beta}$ vs
$(H/M)^{1/\gamma}$]  for (a) $y=0.6$ and (c) $0.8.$  The critical
isotherm, which almost passes through the origin gives an estimate
of respective  $T_C \simeq 222.5$ K and   $241$ K. (b) and (d) :  A
plot of $M|\epsilon|^{-\beta}$   against $H
\epsilon|^{-(\beta+\gamma)}$ show  scaling collapse  for $y=0.6,0.8$
respectively.}
\end{figure}

\section*{An alternative  scaling}

In the main text we  have chosen   the  exponents $\gamma$ and
$\delta$ to  scale independently
\begin{equation}
\gamma \to
\frac{\gamma}{\lambda^{\omega}};
\frac{\delta+1 }{\delta-1}  \to \frac{1}{\lambda^{\kappa}} \frac{\delta+1 }{\delta-1}.
\end{equation}
But, since   magnetic  phase transitions  are  associated with {\it two} independent critical exponents,
one   can vary  any  two independently  and fix  variation of  others  through scaling relations.
In  theoretical studies, $\eta$  and $\nu$ are  natural choices  as they can be obtained from the  two-point
correlation function,
\begin{equation}
\langle S(\vec R ) S(\vec R +\vec  r) \rangle    -  \langle S(\vec R)  \rangle^2
= \frac{e^{- |\vec r |/\xi}}{|\vec r|^{d-2+\eta}},
\end{equation}
where  the correlation length  $\xi\sim    (T_c- T) ^{-\nu}.$  Also, $\eta$   satisfy another
hyper  scaling relation,
\begin{equation}
 2-\eta  =  d  \frac{\delta-1 }{\delta+1} =   \frac{\gamma} {\nu}.
\end{equation}
Thus the  variation we  propose in  the  main text (Eq. (5)),  is equivalent to
\begin{equation}
  2-\eta = \lambda^{\kappa} (2-\eta_0) ~;~ \nu= \frac{\nu_0}{\lambda^{\kappa +\omega}}.
\end{equation}
Other  exponents  can be   derived as  they  are functions  of  $(\eta, \nu),$
\begin{eqnarray}
 \gamma =  \nu (2-\eta) ~;~\alpha = 2-  d\nu    ~;~ \beta  = \frac{\nu}{2} (d-2+\eta) ~;~
 \delta =  \frac{d+2-\eta}{d-2+\eta} \nonumber\
 \end{eqnarray}
when  scaling relations  hold.